\documentclass[3p,times,twocolumn]{elsarticle}
\usepackage{amssymb}
\journal{Journal of Magnetism and Magnetic Materials}

\begin{document}

\begin{frontmatter}

\title{Anisotropic magnetothermoelectric power of ferromagnetic thin films}

\author[1,2]{M. S. Anwar\corref{mycorrespondingauthor}}
\cortext[mycorrespondingauthor]{Corresponding author}
\ead{m.s.anwar@gmail.com}

\author[1,3]{B. Lacoste}

\author[1]{J. Aarts}
\address[1]{Kamerlingh Onnes Laboratory, Leiden University, P.O. Box 9504, 2300 RA Leiden, The Netherlands}
\address[2]{London Centre for Nanotechnology, University College London, WC1H 0AH, United Kingdom}
\address[3]{SPINTEC, UMR CEA/CNRS/UJF/INPG, CEA/INAC, 17 Rue des Martyrs, 38054 Grenoble, France}

\begin{abstract}

In this article, we report the measurements of the magnetothermoelectric power (MTEP) in metallic ferromagnetic thin films of Ni$_{80}$Fe$_{20}$ (Permalloy;  Py), Co and CrO$_{2}$ at temperatures in the range of 100~K to 400~K. In 25~nm thick Py films and 50~nm thick Co films both the anisotropic magnetoresistance (AMR) and MTEP show a relative change in resistance and thermoelectric power (TEP) of the order of 0.2$\%$ when the magnetic field is reversed, and in both cases there is no significant change in AMR or MTEP after the saturation field has been reached. Surprisingly, both Py and Co films have opposite MTEP behaviour although both have the same sign for AMR and TEP. The data on half metallic ferromagnet CrO$_2$ films show a different picture. Films of thickness of 100~nm were grown on TiO$_2$ and on sapphire. The MTEP behavior at low fields shows peaks similar to the AMR in these films, with variations up to 1$\%$. With increasing field both the MR and the MTEP variations keep growing, with MTEP showing relative changes of 1.5$\%$ with the thermal gradient along the $b$-axis and even 20$\%$ with the gradient along the $c$-axis, with an intermediate value of 3$\%$ for the film on sapphire. It appears that the low-field effects are due to the magnetic domain state, and the high-field effects are intrinsic to the electronic structure of CrO$_2$ and intergarian tunnelling magnetoresistance that contributes to MTEP as tunnelling-MTEP. Our results will stimulate the research work in the field of spin dependent thermal transport in ferromagnetic materials to further develop spin-Caloritronics.

\end{abstract}


\end{frontmatter}


\section{Introduction}

Electronic transport in ferromagnets is spin-dependent, which conceive several phenomena including anisotropic magnetoresistance (AMR). In magnetic junctions it comes out as giant magnetoresistance (GMR) and tunnelling magnetoresistance (TMR) those are the backbone of Spintronics. On the other hand, thermoelectric effects are known since nineteenth century, the Seebeck effect in particular is used in thermocouples. In ferromagnets it is predicted that heat transport is also spin-dependent, more precisely that thermoelectric power (TEP) is spin-dependent~\cite{Bauer2012}. Several experiments have been illustrated it and open a new field, the so-called "Spin Caloritronics". Such experiments are for example the Spin-Seebeck Effect (SSE)~\cite{Uchida2008}, Spin Peltier effect~\cite{Flipse2012} TEP in multilayer nanowires~\cite{Gavier2006,Mitdank2012} or thermally induced spin torque in nonlocal lateral spin-valves~\cite{Slachter2010}. Recently an equivalent to TMR for thermoelectric transport was observed, the tunnelling magnetothermoelectric power (TMTEP)~\cite{Liebing2011}. Its relative magnitude was found to be of the same order as TMR. However magnetoresistance and magneto-thermoelectric power (MTEP) are not directly related in theory~\cite{phd-thesis}, the intrinsic conductivity is proportional to the density of states (DOS) at the Fermi level, whereas the intrinsic TEP depends on the derivative of the DOS, through the derivative with respect to energy of the conductivity $\sigma\acute{}$. The TEP is given by~\cite{Mott1958},

\begin{equation}    
	S = -eL_{\circ}T\frac{\sigma\acute{}(\epsilon_{F})}{\sigma(\epsilon_{F})}
	\label{MottsLaw}
\end{equation}

It is known as Mott's Law of TEP~\cite{Jonson1980}. Here, $L_{\circ}$ is the Lorentz number, which is a universal quantity $L_{\circ} = 2.45 \times 10^{-8}~W\Omega K^{-2}$. This expression also illustrates that TEP is linear with $T$ which is true only for diffusive electronic contribution. A $T^3$ dependence term is added if the phonon drag phenomenon is also contributing to the TEP. In a ferromagnetic material the magnon drag phenomenon can contribute along with diffusive and phonon drag. The magnon drag part is dependent on $T^{3/2}$ which makes it difficult to differentiate the both phonon and magnon contributions~\cite{Ziman1964}.  

Note, eq.~\ref{MottsLaw} is valid for a homogeneous conductor having only one kind of carriers and also when variation with respect to energy of the mean free path ($\lambda_{\epsilon}$) and relaxation time ($\tau_{\epsilon}$) are negligible. A ferromagnet has different DOS at the Fermi level for spin up ($N_{\uparrow(\epsilon_F)}$) and spin down electrons ($N_{\downarrow(\epsilon_F)}$) and in this sense Mott's Law can be written as~\cite{Blatt1976},

\begin{equation}    
	S=-eL_{\circ}T\left(\frac{\sigma\acute{}_\uparrow(\epsilon_{F})}{\sigma_\uparrow(\epsilon_{F})}+\frac{\sigma\acute{}_\downarrow(\epsilon_{F})}{\sigma_\downarrow(\epsilon_{F})}\right) = \left(S_{\uparrow} + S_{\downarrow}\right)
	\label{MottsLawSum}
\end{equation}

Still, it is valid in situations when $\lambda_{\epsilon}$ and $\tau_{\epsilon}$ do not vary significantly. But in inhomogeneous systems such as magnetic multilayer, or magnets in a domain state, this will be different. For instance, Piraux {\it et al.}~\cite{Piraux1992} investigated the consequences of electron-magnon scattering in the framework of GMR, which is instructive to mention here. A spin down (up) electron of wave vector $k$ can be scattered into a spin up (down) state with wave vector $k\acute{} = k \pm q$ creation (annihilation) of a magnon with wave vector $q$. The magnon energy $E_q$ will transfer to or from the electron. This results to a maximum scattering rate for spin up electrons with energies below the Fermi level $\epsilon_F$, and/or spin down electrons with energies above. As a result the relaxation rate at $\epsilon_F$ will have different signs and opposite effect for a spin down electron. That results to maximum rate of scattering below and above the Fermi level for spin up and spin down electron respectively. As a result, the magnon scattering contribution to the thermopower is $S^m_{\uparrow(\downarrow)} = \mp {L_\circ}/{k_B}$, and we get,

\begin{equation} 
	S^m =\frac{N_{\uparrow}(\epsilon_F)S^m_{\uparrow} - N_{\downarrow}(\epsilon_F)S^m_{\downarrow}}{N_{\uparrow}(\epsilon_{F}) + N_{\downarrow}(\epsilon_{F})} \label{Piraux}
\end{equation}

Note, this contribution to the thermopower will go to zero for weak ferromagnets where the difference between the DOS for spin-up and spin-down is small. Otherwise, the difference for the different spin channels has consequences for the TEP in F/N multilayers when the magnetic configuration is changed from parallel to antiparallel.

A ferromagnetic material has different Seebeck coefficients (or TEP) for spin up and spin down electrons analogous to spin dependent electrical conductivity. Hence, the spin dependent thermal transport has a relation with the spin dependent electronic transport. Hence we can measure an anisotropic magnetothermoelectric power (AMTEP) by applying an external magnetic field to change the magnetic domains in particular in ferromagnetic thin films. This question has been addressed for thin films of ferromagnetic semiconductor Mn-doped GaAs, where longitudinal and transverse MTEP was measured with in plane applied magnetic field. The effect was found to be related to the AMR and Planner Hall effect (PHE)~\cite{Pu2006}. 

We studied AMTEP in thin films (100~nm thick) of fully spin polarized CrO$_{2}$ ferromagnetic metal. It is strongly related with the magnetization structure, like AMR and we observed an MTEP signal many folds larger than AMR. We also measured the MTEP for partially spin polarized materials, namely Co (50~nm thick) and Py (25~nm thick), for which we measured an MTEP of the same order of magnitude as AMR. The temperature dependent Seebeck coefficient $S$($T$) is strongly dependent on crystalline axis for CrO$_{2}$ films deposited on TiO$_{2}$ and it is almost linear for CrO$_{2}$ deposited on sapphire substrates. Note that all three ferromagnetic thin films used in this study have different thickness. In fact, TEP and resistivity data do not dependent on the thickness of the films. Thus results of these different films can be analysed comparatively. 

\section{Experimentation}

\begin{figure}[h]
	\begin{center}
		\includegraphics[width=8cm]{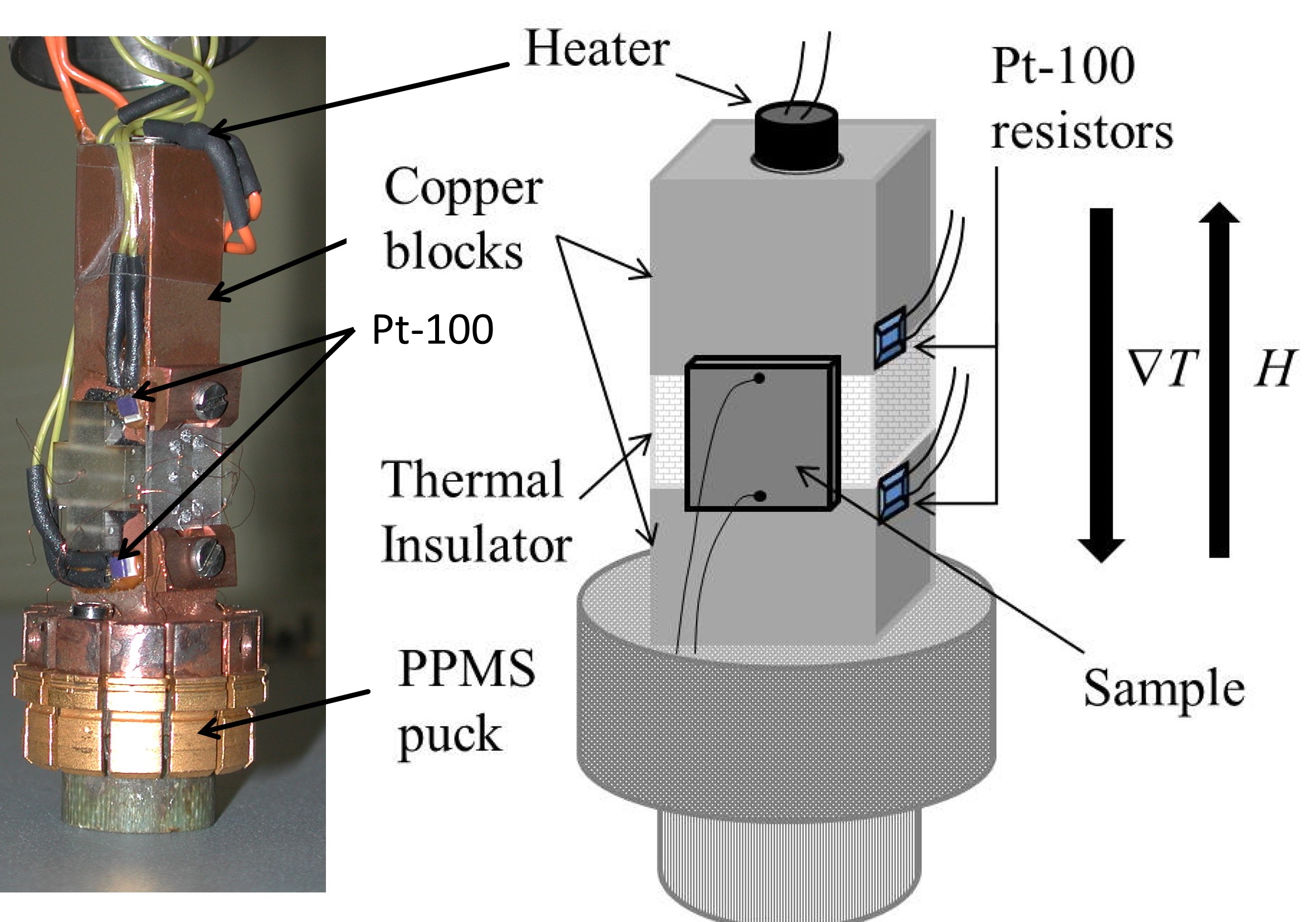}
		\caption{(Left) A photo of thermal transport sample holder built on a PPMS puck and (right) a schematic view. Two copper blocks separated by a thermally insulator plastic, a heater is installed in the upper Cu block and two Pt-100 resistor are being used to measure the temperature difference between the Cu blocks. Whole set up is covered with a stainless steel cap that is not shown in this picture.}\label{sampleholder}
	\end{center}
\end{figure}

The Seebeck coefficient was measured using a home-made sample holder built on a PPMS puck. It consists of two copper blocks (1~cm$^3$) separated by a thermal insulator plastic. The copper has a high thermal conductance so the blocks are at a uniform temperature while a temperature gradient is produced between them. A small heater (maximum power of 5~W) is installed in the upper block. Its temperature is measured with a Pt-100 resistor and controlled with an external temperature controller. The temperature of the lower block is controlled by the set point of the PPMS, but the temperature was separately measured by a second Pt-100 resistor. The whole set-up is covered with a stainless steal cup that isolates the sample holder from radiation loss and it helps to stabilize the temperature gradient. The measurements were done in a relatively low vacuum of 10$^{-2}$~mbar. A photograph and a schematic view of the sample holder is given in Fig.~\ref{sampleholder}

The samples consisted of thin films, mostly on sapphire substrates, with an area of 10 $\times$ 10~mm$^2$. CrO$_{2}$ thin films with thickness of 100~nm were deposited by chemical vapor deposition (CVD) on both isostructural TiO$_{2}$(100) and sapphire (1000) substrates. CrO$_{2}$ film deposits epitaxially on TiO$_{2}$ in the form of rectangular grains aligned along the $c$-axis but on sapphire the grains are aligned with six fold rotational symmetry coming from the hexagonal structure of the substrate, for details see the Ref.~\cite{Anwar2011,Anwar2013}. The Py thin films of thickness of 25~nm were deposited using dc sputtering in a UHV sputtering system, with a base pressure of 10$^{-9}$~mbar, the Co films of thickness of 50~nm were deposited in Z-400 an RF sputtering system with base pressure of 10$^{-6}$~mbar. Both Py and Co were deposited on sapphire substrates because of its better thermal conductivity. Quality of the films were checked by measuring ferromagnetic properties scuh as magnetic loops and magnetoresistance.

The Seebeck coefficient was recorded with reference to copper since Cu wires were connected at both ends of the film via pressed Indium. The potential difference was probed using a Nanovoltmeter (Keithley 2018) in an open circuit geometry $(J = 0)$. A dynamic technique was utilized to measure TEP as function of temperature in which the temperature difference between hot and cold point was always 5~K, while the temperature of the cold point was increased by 10~K in each step. In this way hot point and cold point interchanged in each step between the temperature range of 100 - 400~K~\cite{Compans1989}.

\section{Results}

To check our experimental setup, TEP was measured for thin films with thickness of 100-nm of well characterized metals, namely Cu, and Au, with reference to Cu at room temperature. In principal, it should give a zero TEP on a Cu film, but we measured around 0.5~$\mu$V at temperature difference of 10~K, which gives a TEP of the order of 0.05~$\mu$V/K. The small TEP validates the assumption that we can take Cu as a reference for the TEP measurements and also attest our home-built sample holder. The non-zero voltage appearing on Cu film can be attributed to two factors. Firstly, we used pressed Indium to contact the Cu wires with samples that can also contribute to the TEP. Secondly, the Cu wires may not be connected really with the temperature bath, which can still generate some temperature difference between the voltage pads. TEP on Au thin film gives the values of $-$0.4~$\mu$V/K reference to Cu. The absolute TEP of Au at 300~K is 1.94~$\mu$V/K, whereas for Cu, it is 1.84~$\mu$V/K, which makes TEP of Au in reference to Cu to be $+$0.1~$\mu$V/K. We measure a coefficient of $-$0.4~$\mu$V/K at room temperature. The sign change indicates a role of the In, since the absolute TEP of In is around 1.5~$\mu$V/K~\cite{Caplin1974}, giving a negative TEP in reference to Cu bigger than Au.

\begin{figure}[t]
	\begin{center}
		\includegraphics[width=8cm]{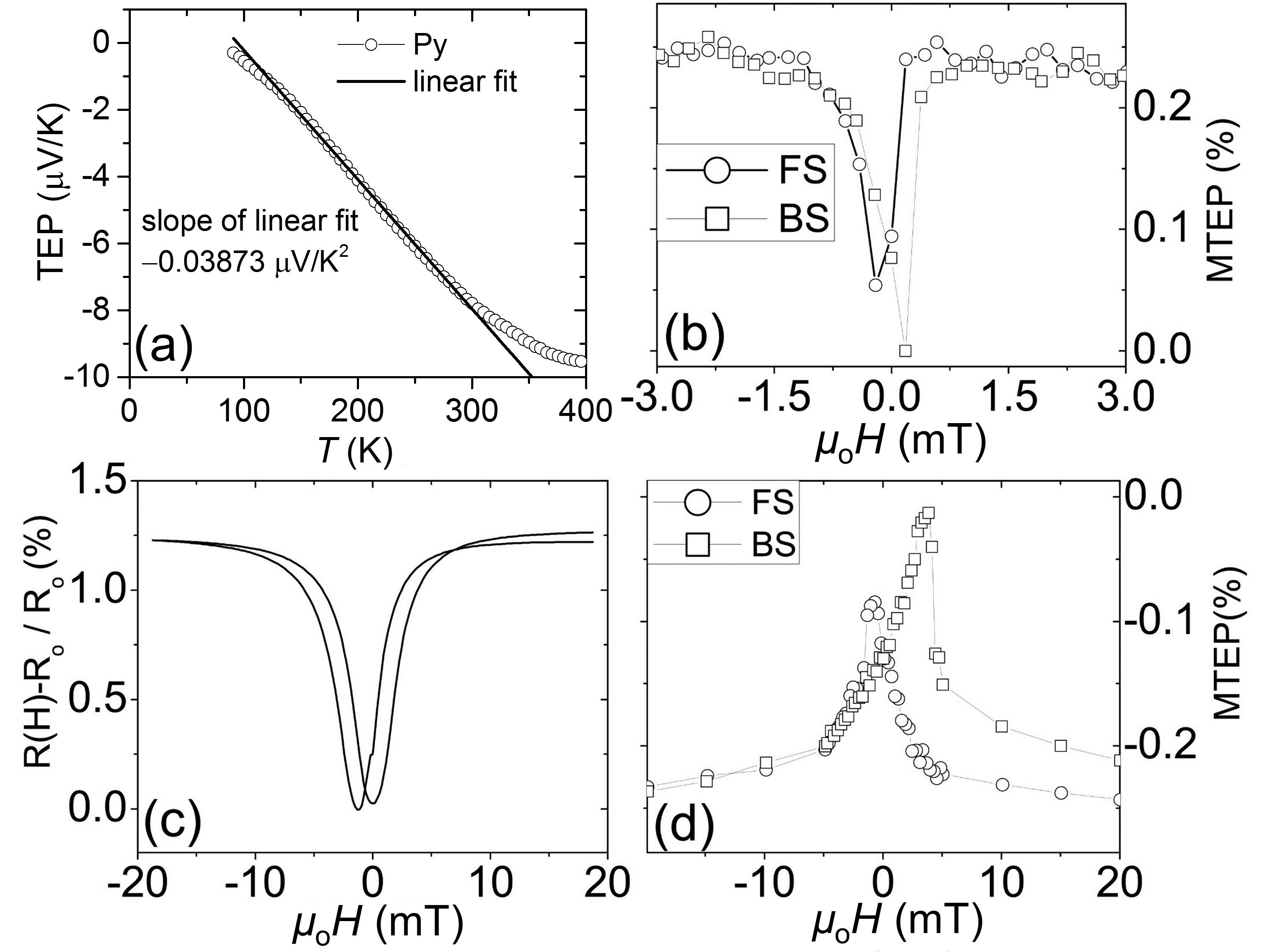}
		\caption{(a) Thermoelectric power (TEP) as a function of temperature for a 25~nm thick Py thin film in the temperature range of 100 - 400~K measured with a constant $\Delta T$ of 5~K. Black solid line inducates a linear fit. (b) MTEP for the same film at $\bar{T}$ = 178~K and $\Delta T$ = 45~K, with the field applied along the temperature gradient. Note that open circles are indicating the Forward Sweep (FS: high to low fields), while open squares are presenting data for Backword Sweep (BS: low to high field). (c) AMR for the same Py film, measured at 4.2~K. (d) MTEP measurements for a 50~nm thick Co thin film measured at $\bar{T}$ = 178~K with $\Delta T$ = 45~K and the field applied along the temperature gradient. Note the difference in sign compared to the data on Py.}\label{SHPyCo}
	\end{center}
\end{figure}

Figure~\ref{SHPyCo}a shows the temperature dependent TEP of a 25~nm thick Py film. It shows roughly linear behavior between 294~K and 117~K with TEP = $-$7.8~$\mu$V/K at 300~K. It is a smaller value than the value reported in the literature, -20~$\mu$V/K~\cite{Uchida2008} or $-$15~$\mu$V/K,~\cite{Uchida2010}. The reasons for these differences can be the effect of oxide layer on these films and/or dimensionality of the samples in the form of thin films~\cite{Schepis1992}. 

Next, a magnetic field $\mu_{\circ}H~\parallel~\Delta T$ is applied and TEP measured. We use the absolute values of the TEP to define the MTEP, as follows:

\begin{eqnarray}
	MTEP = \frac{\left|TEP(H)\right| - \left|TEP_{max/min}\right|}{\left|TEP_{max/min}\right|}
	\label{MTEP}
\end{eqnarray}

where $TEP_{max}$ or $TEP_{min}$ are the maximum or minimum values of TEP corresponding to coervice fields. When TEP is decreasing (increasing) with externally applied field, we used TEP$_{max}$ (TEP$_{min}$) value to calculation the MTEP using eq \ref{MTEP}. As a consequence, a positive relative change means an increase in TEP with field and vice versa. Figure~\ref{SHPyCo}b is presenting TEP as a function of externally applied field in parallel configuration ($\Delta T \parallel \mu_{\circ}H$) on Py at an average temperature $\bar{T}$ = 178~K and $\Delta T$ = 45~K. It is obvious that sharp peaks appear at 0.2~mT and $-$0.2~mT corresponding to the coercive fields. The TEP is higher in the saturation state and lower in a domain state with a relative change of 0.2$\%$. Such a behavior of MTEP is similar to the positive AMR effect (lower resistance in domain state) in Py films, as elaborated in Fig.~\ref{SHPyCo}c. The same effect is observed for a 50~nm thick Co thin film (see Fig.~\ref{SHPyCo}d), where peaks are appearing at $-$0.7~mT and 3.8~mT. It is interesting to note that for Co the TEP is \emph{lower} in saturation state and \emph{higher} in the domain state, opposite to the films of Py, although both Co and Py have the same sign of TEP and AMR~\cite{AMR-Co}. The MTEP for both films saturates just as magnetization or AMR. The hysteretic behaviour of MTEP and peaks around the coercive field reveal the direct connection of MTEP with the magnetization orientation in the films so we termed it anisotropic-MTEP (AMTEP). For both films the MTEP signal at around 200~K is almost the same in magnitude.

\begin{figure}[t]
	\begin{center}
		\includegraphics[width=8cm]{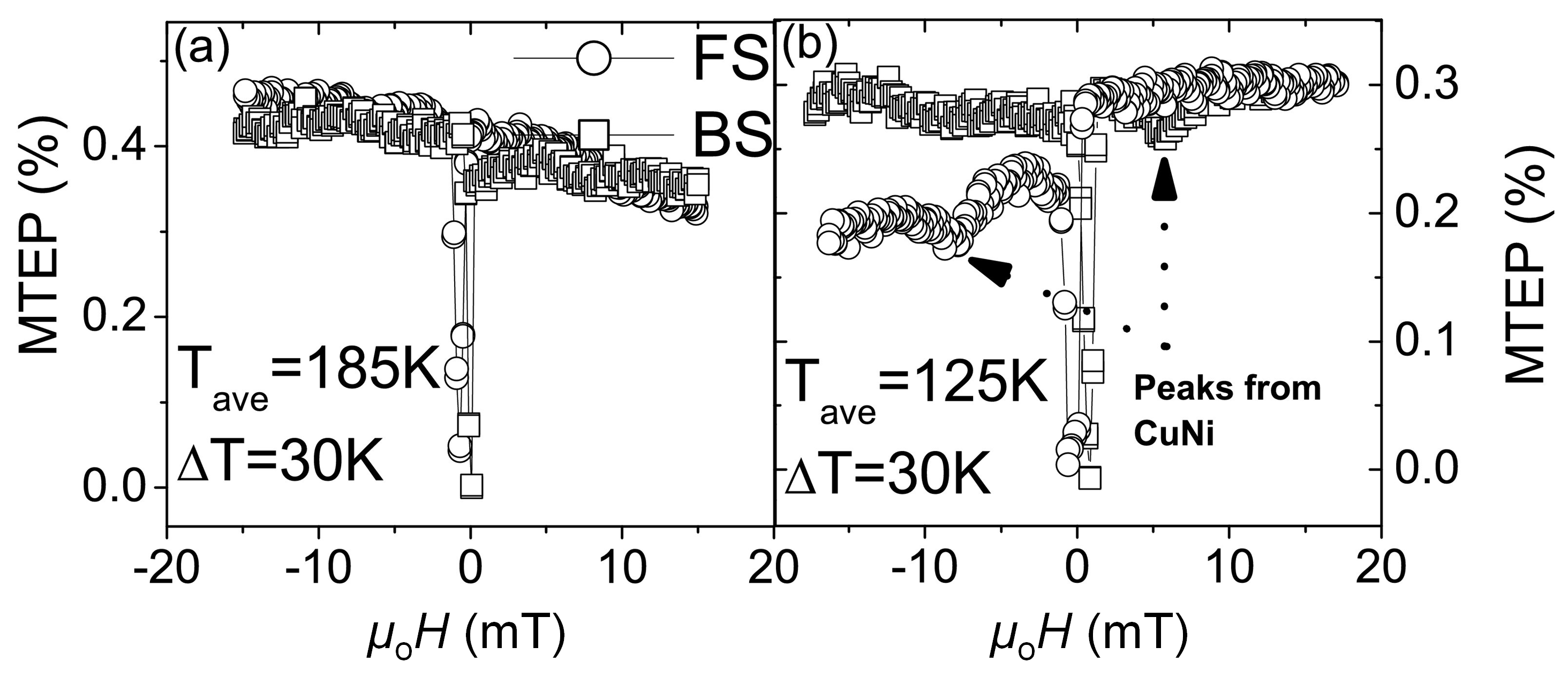}
		\caption{MTEP for a multilayer CuNi(50)/Py(25) film deposited on a sapphire substrate. (a) At $\bar{T}$ = 185~K, which is higher than Curie temperature $T_\textrm{Curie}$ of CuNi (150~K), when there are only two strong peaks corresponding to coercive filed of Py. (b) At lower temperature $\bar{T}$ = 125~K, two more peaks appear corresponding to CuNi.}\label{Shmulti}
	\end{center}
\end{figure}

The effect of magnetization on MTEP is also studied using a sample consisting of a bilayer ferromagnet of CuNi(50~nm)/Py(25~nm) deposited on a sapphire substrate. Cu$_{41}$Ni$_{59}$ is a weak ferromagnet with a $T_\textrm{Curie}$ of about 150~K. Figure~\ref{Shmulti}a shows the data at $\bar{T}$ = 185~K and $\Delta T$ = 30~K, where two central sharp peaks appear. These peaks correspond to the Py, because of low coercive field and sharp switching, equivalent to thin film with Py only. The data recorded at $\bar{T}$ = 125~K, which is lower than the $T_\textrm{Curie}$ of CuNi, shows two more rather wide peaks besides the main central peaks of Py (see Fig.~\ref{Shmulti}b). These additional peaks appearing around 7~mT correspond to the coercive field of CuNi. It demonstrates a strong relation of this effect to magnetization and purely a magnetic domain effect like AMR.

\begin{figure}[h]
	\begin{center}
		\includegraphics[width=8cm]{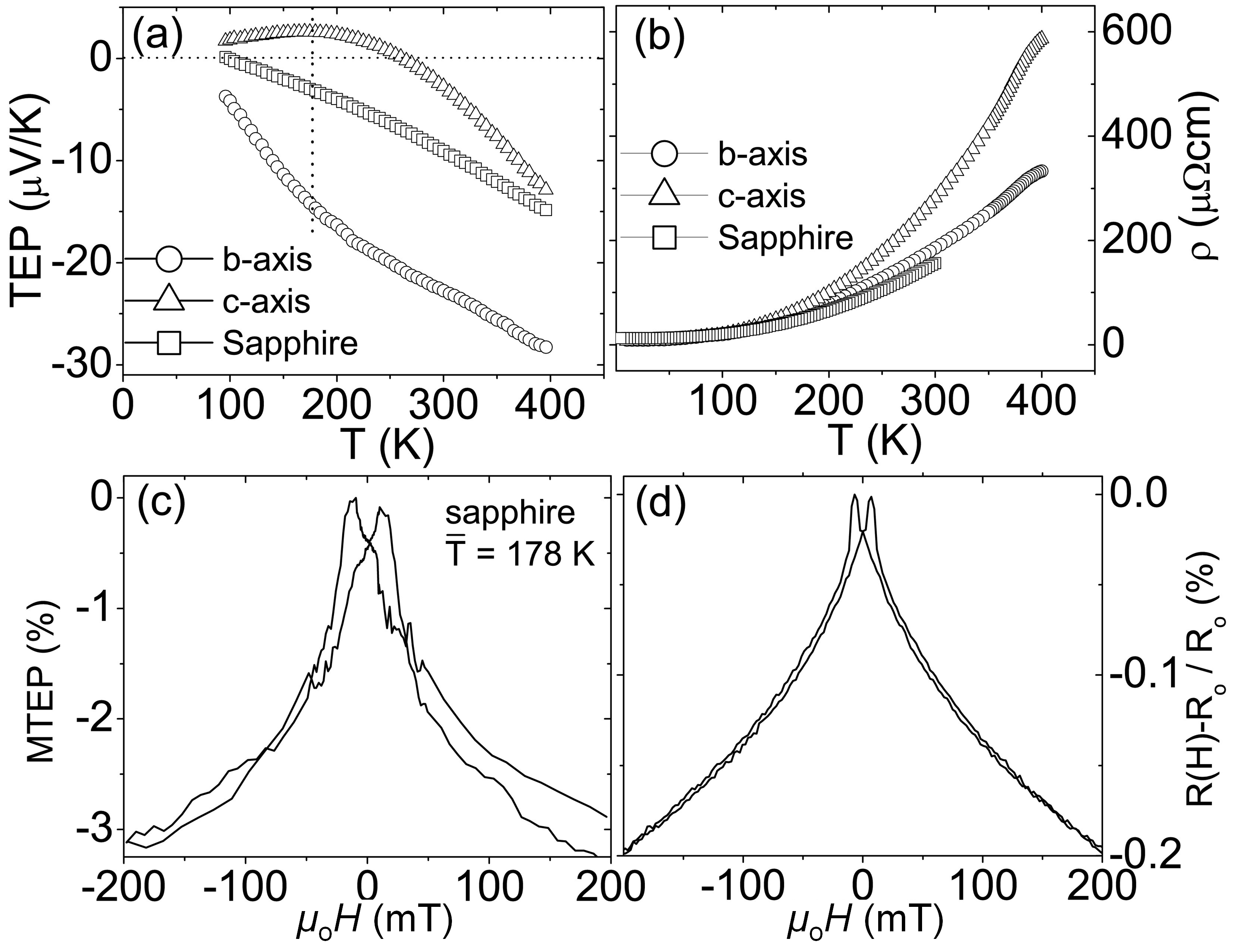}
		\caption{(a) Temperature dependent TEP measured in the temperature range of 100 - 400~K on 100~nm thick CrO$_{2}$ thin films deposited on sapphire ($\Box$) and TiO$_{2}$ substrate along the crystallographic $c$-axis ($\triangle$) and the $b$-axis ($\bigcirc$). TEP changes the sign at 265~K along the $c$-axis. (b) $\rho(T)$ for both films deposited on sapphire and TiO$_2$ substrates. (c) TEP as a function of externally applied field in parallel configuration for CrO$_2$ film deposited on sapphire at $\bar{T}$ = 178~K and $\Delta T$ = 45~K. (d) MR probed on same film at 4.2~K in parallel configuration ($\mu_{\circ}H \parallel I$).}\label{Sthsapphire}
	\end{center}
\end{figure}

Figure~\ref{Sthsapphire}a shows the temperature dependent TEP for CrO$_{2}$ thin films deposited on sapphire and TiO$_{2}$ substrates, with the field and temperature gradient along the $c$-axis (easy axis) and the $b$-axis (hard axis). A room temperature value of $-$9~$\mu$V/K is measured for a CrO$_{2}$ film deposited on a sapphire substrate. Also the TEP decreases linearly with decreasing temperature and approaches zero at 100~K. However for CrO$_{2}$ film deposited on TiO$_{2}$, the TEP has a nonlinear dependent with respect to temperature for applied field along both the $c$-axis and the $b$-axis. TEP along the $c$-axis changes sign at 265~K and has a room temperature value of $-$3~$\mu$V/K. Along the $b$-axis, CrO$_{2}$ has a negative TEP for the whole temperature range, with room temperature value of $-$23~$\mu$V/K, which is quite similar to the literature value of $-$25~$\mu$V/K~\cite{Chapin1965}, where the crystallographic axis is not mentioned.  

Figure~\ref{Sthsapphire}c presents MTEP for a 100~nm thick CrO$_2$ thin film deposited on a sapphire substrate at $\bar{T}$ = 178~K and $\Delta T$ = 45~K and in parallel configuration. TEP is maximum in the domain state and starts to decrease with the increase in field. Two strong peaks at 10~mT appear. They correspond to the peaks at coercive field also visible in AMR as shown in the Fig.~\ref{Sthsapphire}d. For higher magnetic field, more than 20~mT, MTEP and AMR show the same field dependence in CrO$_2$ thin films. The relative change between maximum and minimum TEP at 200~K is about $-$3$\%$, which is 15 times larger than AMR signal that even recorded at 4.2~K.

\begin{figure}[h]
	\begin{center}
		\includegraphics[width=8cm]{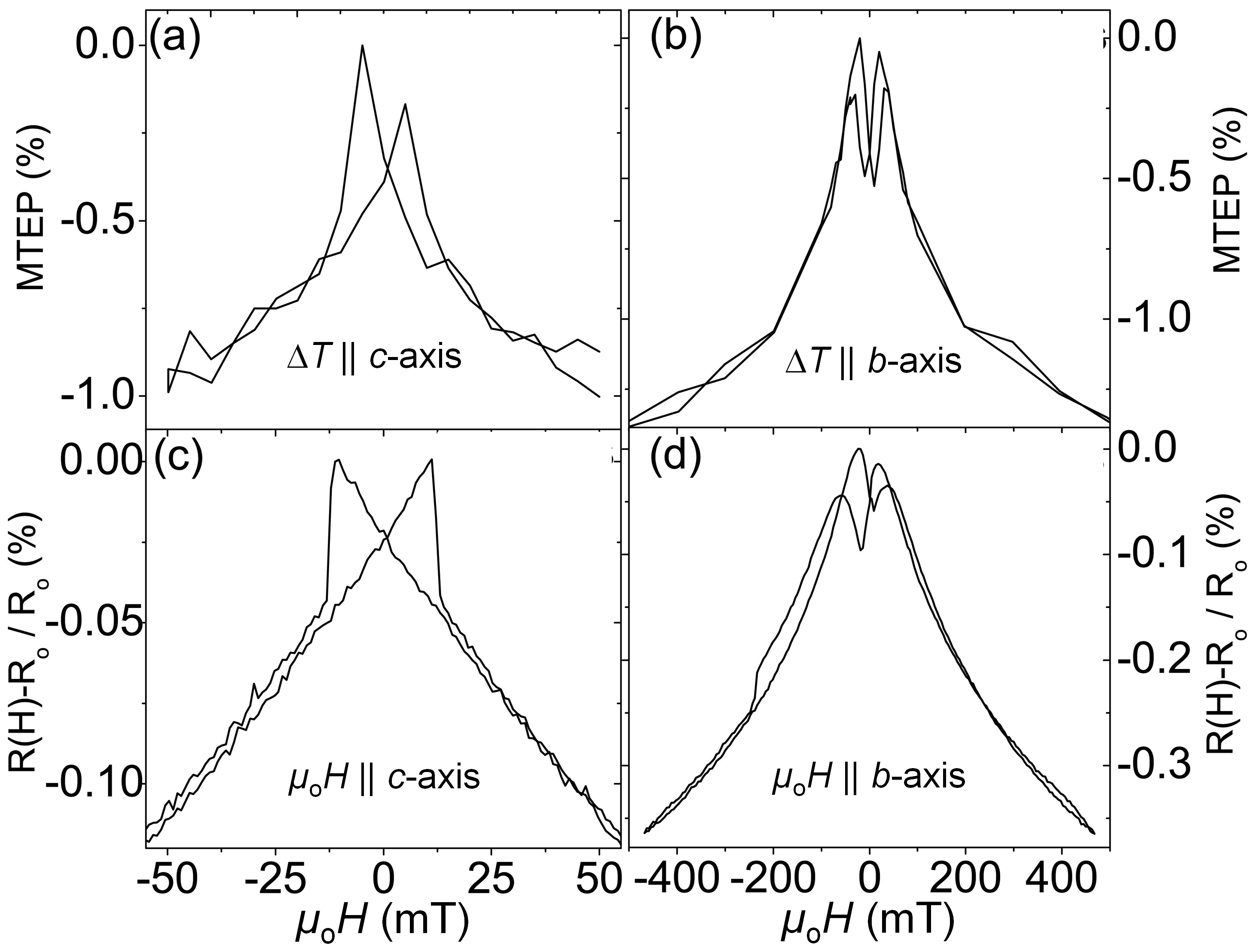}
		\caption{(a) MTEP measured at $\bar{T}$ = 178~K and $\Delta T$ = 45~K, along the $c$-axis of a 100~nm thick CrO$_2$ thin film deposited on TiO$_2$ substrate. (b) MTEP along the $b$-axis. (c) AMR measurements with a dc current of 100~$\mu$A on same film along $c$-axis and (d) along the $b$-axis. The peaks corresponding the coercive field are identical for both cases AMR and MTEP.}\label{Shtio2}
	\end{center}
\end{figure}

Figure~\ref{Shtio2} shows the data of MTEP and AMR data along both in-plane axes (the $c$-axis and the $b$-axis) for a 100~nm thick CrO$_{2}$ thin film deposited on a TiO$_{2}$ substrate. For MTEP measurements the magnetic field $\mu_{\circ}H$ and $\Delta T$ are parallel to each other, as are field and current (100~$\mu$A) for AMR measurements. Along the $c$-axis the MTEP data show two peaks at the coercive field. CrO$_2$ has negative relative change in TEP with field so in the domain state the TEP is higher than in the saturation state. The peaks at coercive field are very sharp in both AMTEP and AMR when the field is applied along the $c$-axis, whereas along the $b$-axis the peaks are less pronounced because of the smooth changes in the magnetic domains along the hard axis. These graphs show a very close correlation with MTEP and AMR or MR, which are both directly sensing the magnetization structure of the samples.

\begin{figure}[h]
	\begin{center}
		\includegraphics[width=8cm]{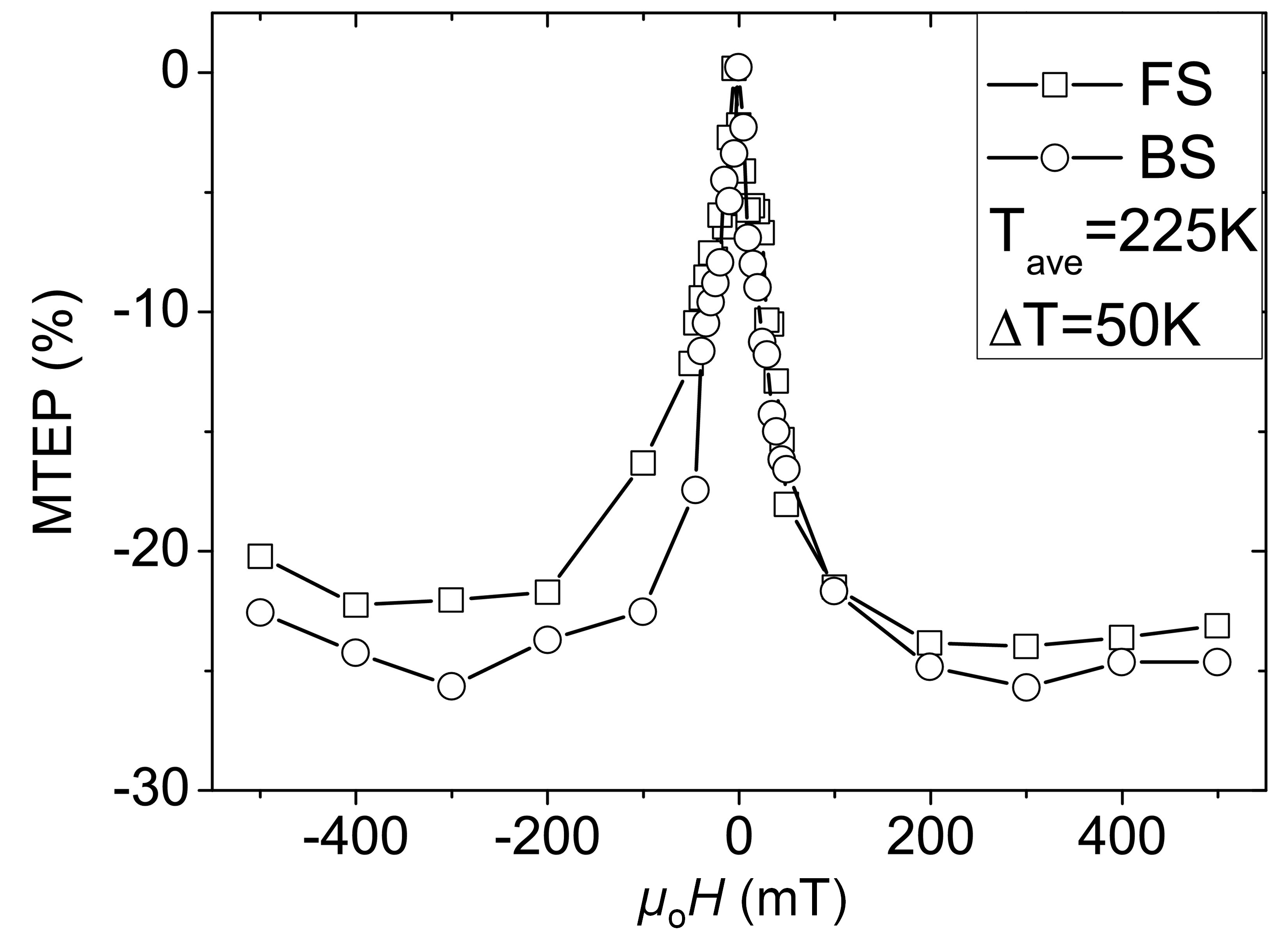}
		\caption{ MTEP measurements along the $c$-axis of a 100~nm thick CrO$_{2}$ film deposited on a TiO$_{2}$ substrate at $\bar{T}$ = 225~K and $\Delta T$ = 50~K with hot point at 250~K. Maximum relative change in MTEP is of the order of 20$\%$.}\label{Shhuge}
	\end{center}
\end{figure}

The relative change in MTEP along the $c$-axis is already 1$\%$ just in 50~mT at around 200~K. MTEP to higher fields shows more than 20$\%$ relative change, see Fig.~\ref{Shhuge}. Note that highest relative change in MTEP for CrO$_2$ occurs just close to the temperature where sign of TEP is changing. It may be one of the main reasons of this rather huge relative change in MTEP. Along the $b$-axis, the relative change in MTEP is 5 times larger than the MR in the range of 50~mT. MTEP value is always much larger than MR regardless of the substrate used to grow the thin films of CrO$_2$. In Fig~\ref{Shhuge}, the saturation of MTEP at high fields, above 100~mT, is also obvious.         

\section{Discussion}
\begin{figure}[h]
	\begin{center}
		\includegraphics[width=\linewidth]{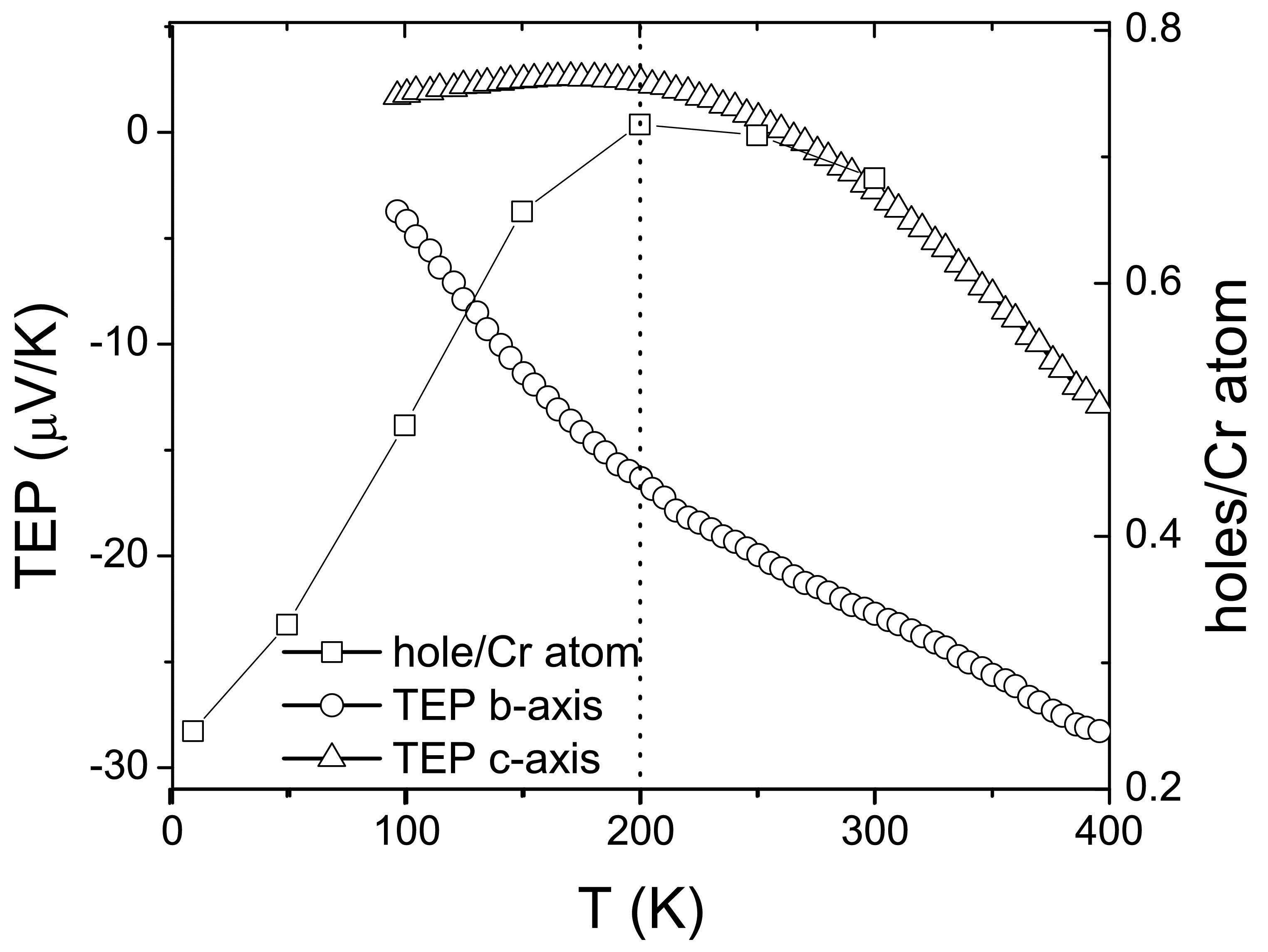}
		\caption{Temperature dependent TEP and carrier concentration of a 100~nm thick CrO$_{2}$ film deposited on a TiO$_{2}$ substrate. An obvious change in the slope of the TEP occurs around 200~K where the carrier concentration starts to decrease. Note that the TEP data presented here to compare with carrier concentration is already given in Fig. 4.}\label{TEPcarrier}
	\end{center}
\end{figure}

Two aspects of the data can be discussed, the temperature dependence of TEP in CrO$_2$ and Py, and the field dependence found in CrO$_2$, Py and Co. Regarding the temperature dependence, the TEP comes from an electronic contribution, as discussed above, but also, the so-called phonon drag and magnon drag can give a contribution to the TEP. In phonon (magnon) drag, the phonon (magnon) bath is pushed out of equilibrium by scattering with electrons. The resulting net phonon (magnon) momentum yields a drift which also transports heat. Electron-phonon scattering is generally predominant around the Debye temperature, the electron scattering is predominant for lower temperatures and phonon-phonon scattering for higher temperature. Electron-magnon scattering plays also an important role in ferromagnetic materials, as addressed in this paper. In normal metals, the TEP shows a linear behavior versus temperature below the Debye temperature where the electronic contribution is predominant. Around the Debye temperature, the electron-phonon scattering plays an important role, and the linear approximation is generally not valid anymore. The TEP measured in Py (Fig.~\ref{SHPyCo}a) shows an almost linear behavior between 100~K and 300~K indicating that the electronic contribution dominates. For CrO$_2$, TEP for films on TiO$_2$ exhibits non-linear temperature variations, as well as a difference between the $c$-axis and the $b$-axis. What is remarkable is the change in behavior around 200~K. Along the $b$-axis, TEP versus temperature curve shows a kink at around 200~K. Along the $c$-axis, the TEP slope vanishes around 200~K. We compare this behavior with that of the carrier concentration $n$($T$) found via Hall Effect measurements Fig.~\ref{TEPcarrier}. The TEP change of slope at around 200~K corresponds to a maximum in $n$($T$). The films deposited on sapphire show a linear relation like Py. For these films the rectangular grains are randomly oriented that might give a combined effect along both the $c$- and $b$-axes, thus shadowing the previous effects.

Next, we focus on the variation of TEP with externally applied magnetic field. MTEP measurements present in this paper are all obtained with a temperature difference of 45~K. Notice that the MTEP data for smaller temperature differences, down to 5~K difference, give similar results but with smaller AMTEP ratio, around $1\%$. However for small temperature differences the noise level is more important, that is why only results for 45~K temperature difference are shown here. The voltage difference measured is not exactly proportional to $S$($T$) for large temperature difference, but instead $V=\int_{T1}^{T2} S$($T$)$dT$. This does not change qualitative conclusions. In particular, MTEP, like MR, gives an insight of the magnetic state of the sample. Both curves show dramatic changes at the coercive field, and for large applied fields in absolute value, MTEP is linear, corresponding to a saturated magnetic state.

For Py (Fig.~\ref{SHPyCo}b-c), MTEP and MR curves show a very similar dependence. In the domain state the AMR shows a lower resistance (field $\parallel$ current), when MTEP shows a lower voltage (field $\parallel$ thermal gradient) than in the saturated state. Note that the AMR was measured at 4.2~K, the effect at 200~K is much smaller. MTEP for Co shows similar MTEP behavior, except that the thermal voltage is now increased in the domain state as compared to the saturated state. It is well known that the AMR has the same sign for Co and Py~\cite{AMR-Co,Fert1999}, which therefore also might have been expected for MTEP. It may be explained by the fact that the difference of TEP for majority and minority spin electrons is not similar to the difference of resistance of majority and minority spin electrons.

For CrO$_2$, as shown in Fig.~\ref{Shtio2}, the MTEP behavior faithfully mimics the MR, both for the situations $H~\parallel~c$ (easy axis) and $\mu_{\circ}H~\parallel~b$. It is interesting to note that the MTEP variation is large, of the order of 1$\%$, which is both significantly larger than the MTEP effect in Py, and than the AMR effects in general. The variations of MTEP is such that the thermopower is enhanced in the domain state. 

Since the conductivity $\sigma$ is given by $\sigma ={e^{2}\lambda_{\epsilon}^{2}}N_{\epsilon_{F}}/{\tau_{\epsilon}}$, where $N_{\epsilon_{F}}$ is the DOS at the Fermi level, injecting this relation in eq.~\ref{MottsLaw} Mott's law becomes~:

\begin{eqnarray}  
	S = -eL_{o}T\left\{\frac{N\acute{}(\epsilon_{F})}{N(\epsilon_{F})}+2\frac{\lambda\acute{}_{\epsilon}}{\lambda_{\epsilon}}-\frac{\tau_{\epsilon}\acute{}}{\tau_{\epsilon}}\right\}
	\label{derivativeconduc}
\end{eqnarray}

In a ferromagnet the mean free path and relaxation time are spin dependent. In particular, MTEP variation may follow from the relaxation time energy dependence term $\frac{\tau^{'}_{\epsilon}}{\tau_{\epsilon}}$ in eq.~\ref{derivativeconduc}, but this is not yet completely understood. At high magnetic field, MTEP variations are also similar to MR, but much larger in amplitude. For the sapphire-based film the MTEP change at $\bar{T}$ = 178~K is 3$\%$ between 0~mT and 200~mT, for the TiO$_2$-based film ($\Delta T \parallel c$) is 1$\%$ up to 50~mT at the same temperature. There is not yet much to connect to the experimental or theoretical literature. The comparison of Py and CrO$_2$ indicated that the high spin polarization of the CrO$_2$ is connected to the strong influence of the magnetic scattering on the heat transport. The $c$-axis TEP shows a sign change around 250~K, and the large variation of the MTEP close to this temperature is possibly connected to this sign change. However, the AMTEP is a giant effect compared to AMR for CrO$_2$. Somehow it might be connected to the morphology of the films. The granular nature of CrO$_2$ thin films enhances the contributions of inter-grain tunelling magnetoresistance in MR~\cite{Konig2007}. In this sense TMTEP effect~\cite{Liebing2011} is also contributing to the giant MTEP effect for CrO$_2$ thin films. 

\section*{Conclusions}

The AMTEP was investigated in various ferromagnetic thin films, of Ni$_{80}$Fe$_{20}$ (Permalloy;  Py), Co and CrO$_{2}$ at temperatures in the range of 100~K to 400~K. TEP of Py and Co films depend linearly on the temperature, like CrO$_2$ films deposited on sapphire. For CrO$_2$ films deposited on TiO$_2$ substrate, TEP is nonlinear with a sign change at 265~K along the $c$-axis. The variation of the TEP with temperature appears to be linked to the variation of the carrier concentration. Regarding the MTEP, measured by applying a magnetic field, it is closely linked to the MR because they both describe the magnetic state of the samples. Partially spin polarized thin films of Py and Co show an opposite sign of variation between domain state and saturated state, which is somewhat surprising. CrO$_2$ films, deposited both on sapphire and on TiO$_2$, show large relative changes of the MTEP of the order of 1$\%$, larger than MR changes, of $0.1\%$ at very low temperature. Along the $c$-axis, the voltage variation even reaches 20$\%$ at a temperature close to a sign change of the TEP, but for large temperature difference where the voltage is not exactly proportional to the TEP. Our work will stimulate the theoretical work to understand further the spin dependent thermal transport in ferromagnetic materials.

\section*{Acknowledgements}
We are thankful to Santiago Serrano-Guisan and Shingo Yonezawa for fruitful discussions.

\end{document}